\documentclass[prl,twocolumn,superscriptaddress,showpacs,floatfix,longbibliography]{revtex4-2}
\usepackage{mathrsfs,braket}
\usepackage{amssymb, amsbsy, amsmath, latexsym, dsfont, array, layout, graphicx,mathrsfs,color, bm}
\usepackage[unicode=true,breaklinks=true,colorlinks=true]{hyperref}

\hypersetup{citecolor=blue,urlcolor=blue}

\begin{document}

\title{Non-Hermitian sensing in the absence of exceptional points}

\author{Lei Xiao} \thanks{These authors contributed equally to this work.}
\affiliation{School of Physics, Southeast University, Nanjing 211189, China}

\author{Yaoming Chu} \thanks{These authors contributed equally to this work.}
\affiliation{School of Physics, International Joint Laboratory on Quantum Sensing and Quantum Metrology,
Hubei Key Laboratory of Gravitation and Quantum Physics, Institute for Quantum Science and Engineering,
Wuhan National High Magnetic Field Center, Huazhong University of Science and Technology, Wuhan 430074, China}

\author{Quan Lin}
\affiliation{School of Physics, Southeast University, Nanjing 211189, China}

\author{Haiqing Lin}
\affiliation{School of Physics, Zhejiang University, Hangzhou 310030, China}

\author{Wei Yi}\email{wyiz@ustc.edu.cn}
\affiliation{CAS Key Laboratory of Quantum Information, University of Science and Technology of China, Hefei 230026, China}
\affiliation{CAS Center For Excellence in Quantum Information and Quantum Physics, Hefei 230026, China}

\author{Jianming Cai}\email{jianmingcai@hust.edu.cn}
\affiliation{School of Physics, International Joint Laboratory on Quantum Sensing and Quantum Metrology,
Hubei Key Laboratory of Gravitation and Quantum Physics, Institute for Quantum Science and Engineering,
Wuhan National High Magnetic Field Center, Huazhong University of Science and Technology, Wuhan 430074, China}

\author{Peng Xue}\email{gnep.eux@gmail.com}
\affiliation{School of Physics, Southeast University, Nanjing 211189, China}

\begin{abstract}
Open systems possess unique potentials in high-precision sensing, yet the majority of previous studies rely on the spectral singularities known as exceptional points. Here we theoretically propose and experimentally demonstrate universal non-Hermitian sensing in the absence of exceptional points. The scheme makes use of the intrinsic sensitivity of a non-Hermitian probe to weak external fields, which can be understood as the direct consequence of non-Hermiticity. We confirm the basic mechanism by simulating the sensor-field dynamics using photon interferometry, and, as a concrete example, demonstrate the enhanced sensing of signals encoded in the setting angle of a wave plate. While the sensitivity of the probe is ultimately limited by the measurement noise, we find the non-Hermitian sensor showing superior performance under background noises that cannot be suppressed through repetitive measurements. Our experiment opens the avenue of enhanced sensing without exceptional points, complementing existing efforts aimed at harnessing the unique features of open systems.
\end{abstract}

\maketitle

\textit{Introduction.---}High-precision sensing plays an increasingly important role in modern science, and much effort has been invested in designing novel sensing schemes and next-generation sensors~\cite{T1986,DRC17}. A promising route is to exploit the unique sensitivity of open systems to external perturbations, of which an outstanding example is exceptional-point (EP)-based sensing~\cite{W14,W16,LZO+16,RHH+17,S17,L18,MGK+18,CJL19,ZSH+19,DPD19,MA19,W20,W20PR,BB20,W20PRA,WZL+21,KS22,ASF23,WL23,Ashida2020,Luo2022,Bao2021}. While the dynamics of open systems can be described by non-Hermitian effective Hamiltonians, their complex eigenspectra exhibit exotic degeneracies at the branch-point singularities known as the EPs~\cite{B04,H04,H12}. A system can exhibit fascinating properties near an EP, where, importantly, the eigenspectra become highly susceptible to weak perturbations, a feature that has been extensively researched for EP-enhanced sensing. Over the past decades, proof-of-principle EP-enhanced sensors have been demonstrated using classical electromagnetic or acoustic waves~\cite{FSA15,HHW+17,CKS+17,ZPO+18,LLS+19,XLK+19,NAJ+19,PNC+20,KCE+22,JZL+22,LLX+23}, and in quantum open systems of trapped ions~\cite{DSZ+21}, solid spins~\cite{WLG+19}, or single photons~\cite{YMT+20}. These prototype sensors and their underlying protocols vitally depend on the presence of EPs, necessitating specific designs of non-Hermitian effective Hamiltonians with fine-tuned parameters.

In this work, we propose and demonstrate non-Hermitian sensing in the absence of EPs. The scheme relies on the non-Hermiticity-enhanced susceptibility of physical observables to weak perturbations, which, distinct from the EP-based sensitivity, is a generic feature of non-Hermitian systems, making the sensing scheme universal~\cite{CLL+20, LC18, LLW+23}. Using a cyclic photonic interferometer, we experimentally simulate the discrete-time dynamics of the probe-field system with tunable non-Hermiticity. We observe enhanced sensitivity under a non-Hermitian setting, thus confirming the feasibility of our scheme. To provide a prototype scenario for application, we then demonstrate the sensing of signals encoded in the setting angle of a wave plate. We also theoretically analyze and experimentally confirm the superiority of the non-Hermitian sensing scheme---it outperforms its Hermitian counterpart in the presence of background noises that do not average out through repetitive measurements.

\begin{figure}
\centering
\includegraphics[width=0.5\textwidth]{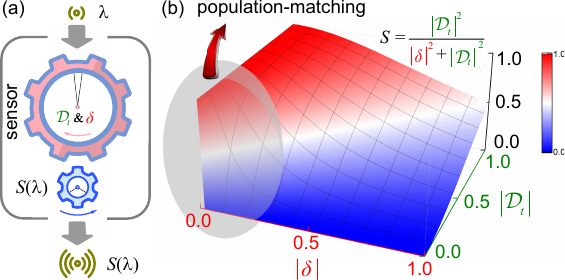}
\caption{Mechanism of the non-Hermiticity-enhanced sensing.
(a) Schematic of the workflow. Within the sensor's range, a tiny variation in an external parameter $\lambda$ leads to small changes in the sensor, parameterized by $\mathcal{D}_t$ and $\delta$.
These changes further give rise to significant variations in a detectable $S(\lambda)$.
(b) A qubit sensor is initialized in the basis state $|0\rangle$, close to an eigenstate (denoted as $|\phi_+\rangle$) of its non-Hermitian Hamiltonian, with $|\phi_+\rangle\sim|0\rangle+\delta|1\rangle$ and $|\delta|\ll 1$ is a parameter of the non-Hermitian sensor.
Under a weak perturbation, the sensor generically evolves into the state $|\psi(t)\rangle\sim \mathcal{D}_t|0\rangle+\delta|1\rangle$.
Generally, when the parameter $\mathcal{D}_t$ is in the same order as $\delta$, with $|\mathcal{D}_t| \sim |\delta|\ll 1$, the final population in state $|0\rangle$ (denoted as $S$) exhibits a sensitive reliance on the small parameters $|\delta|$ or $\mathcal{D}_t$, thus offering a practical probe for these parameters.
}
\label{theory}
\end{figure}

\textit{Theoretical framework.---}We consider a generic non-Hermitian qubit sensor perturbed by a weak external field, with the overall Hamiltonian given by $H=\mathcal{H}+\lambda V$, and $\lambda$ denotes the small parameter to be estimated and  $V=\sigma_x$. In a properly chosen orthonormal basis denoted by $\{|0\rangle,|1\rangle\}$, the Hamiltonian of the bare non-Hermitian sensor can be generally expressed as (up to an arbitrary constant)
\begin{equation}
\mathcal{H}=\frac{ 2 \mathcal{E}}{1+a}\begin{pmatrix}{1}&{\frac{a}{\delta}}\\{\delta} & {a}\end{pmatrix}
\end{equation}
with $|a|\leq 1$. Its two eigenstates are given by $|\phi_+\rangle\sim|0\rangle+\delta|1\rangle$ and $|\phi_-\rangle\sim|0\rangle-(\delta/a)|1\rangle$, respectively, with an associated energy splitting $2 \mathcal{E}$.
%
Importantly, our sensing scheme assumes the condition $|\delta|\ll 1$, which is always possible by a proper choice of the basis. Since the requirement of $|\mathcal{E}|\to 0$ is relaxed, the non-Hermitian sensor under study is independent of the properties of any EPs. The working parameters selected for our experiments below are all far away from any EP, as characterized by the Frobenius norm distance~\cite{HJ12,supp}.


We first focus on the unperturbed system dynamics governed by $\mathcal{H}$ with the sensor initialized in the basis state, $|\psi(0)\rangle=|0\rangle$.
The time-evolved state can then be written as
\begin{equation}
\label{Eq:EvolvedState}
|\psi(t)\rangle=e^{-i\mathcal{H}t}|\psi(0)\rangle=\frac{1-e^{i 2 \mathcal{E} t}}{1+a}\left(\mathcal{D}_t|0\rangle +\delta |1\rangle\right),
\end{equation}
where $2\mathcal{D}_t = (1-a)+i(1+a) \cot \mathcal{E} t $.
The population in the state $|0\rangle$ is straightforwardly determined by
\begin{equation}
\label{Eq:population}
S(t)=\frac{|\mathcal{D}_t|^2}{|\delta|^2+|\mathcal{D}_t|^2}.
\end{equation}

Now, consider the time evolution under $H$, where an additional weak field is present. The parameters of the time-evolved state $\{\delta,a,\mathcal{E} \} $ all become $\lambda$-dependent, so is the normalized population in $|0\rangle$, which we denote as $S(\lambda,t)$. Its sensitivity to $\lambda$ is captured by the susceptibility $\chi_{\lambda}=\partial_\lambda S(\lambda,t)$, which can be decomposed into
\begin{equation}
\chi_\lambda=\chi_{|\delta|} \partial_\lambda |\delta|+\chi_{|\mathcal{D}|} \partial_\lambda |\mathcal{D}_t|,
\label{Eq:susceptibility}
\end{equation}
where $\chi_{|\delta|}$ and $\chi_{|\mathcal{D}|}$ are responses of $S$ to variations in the parameters $|\delta|$ and $\mathcal{D}$, respectively.

As illustrated by Fig.~\ref{theory}, in the region $|\mathcal{D}_t|\sim |\delta|\ll 1$, $S(\lambda)$ shows fast variations with respect to $|\delta|$ and $|\mathcal{D}_t|$, thus also features a highly sensitive response to $\lambda$. Specifically, in the region with $|\mathcal{D}_t|\simeq |\delta|\ll 1$, the populations in the states $|0\rangle$ and $|1\rangle$ are comparable to each other. Due to the smallness of $\delta$ and $\mathcal{D}_t$, a weak perturbation (characterized by $\lambda$) could take the non-Hermitian sensor out of this region, causing a significant change in the signal.

The above population-matching condition (namely $|\mathcal{D}_t|\sim |\delta|\ll 1$) is achieved by exploiting the generic non-Hermiticity of the sensor, inherently facilitated by the non-orthogonal eigenstates and complex eigenvalues. For example, the non-orthogonality of eigenstates in a pseudo-Hermitian system (where $\mathcal{E}$ is real) allows $a$ to be of the order unity ($a\approx 1$), which straightforwardly yields $\mathcal{D}_t\approx i\cot \mathcal{E} t $~\cite{CLL+20}. When the evolution time is close to $\mathcal{E} t\simeq \frac{\pi}{2} $, $|\mathcal{D}_t|$ could be on the order of $|\delta|$, satisfying the population-matching condition. More general non-Hermitian sensing configurations with gain and loss beyond the pseudo-Hermitian case can also be established based on the population-matching condition~\cite{supp}.

The observations above form the basis of our non-Hermiticity-enhanced sensing scheme, which is fundamentally different from the previously reported EP-enhanced sensing~\cite{W14,W16}. While the latter typically relies on the sensitivity of eigenstates and eigenvalues close to the EP, our scheme exploits the non-Hermitian sensing region illustrated in Fig.~\ref{theory}, which is emergent from the non-Hermitian dynamics starting from a specific initial state in the close proximity to an eigenstate of the unperturbed system.

However, the above discussion does not hold for a Hermitian system, where a unitary time evolution gives $2\mathcal{D}_t\approx 1+i \cot \mathcal{E} t $ (where $a=\delta^2\ll 1$), and thus $2|\mathcal{D}_t|\geq 1$. More specifically, we consider an exemplary Hermitian sensor, where the full Hamiltonian is given by $H_s=\lambda\sigma_x$. The time-evolved state follows the form of $|\psi(t)\rangle=e^{-i\lambda\sigma_x t}|\psi(0)\rangle$.
Hence, the normalized population in state $|0\rangle$ is
\begin{equation}
S(\lambda,t)=\frac{1}{2}\left(1+\cos 2\lambda t\right).
\label{hsens}
\end{equation}
The advantage of the non-Hermitian sensor is clearly revealed by comparing the susceptibilities defined in Eqs.~\eqref{Eq:susceptibility} and \eqref{hsens}, respectively, the latter being a smooth-varying sinusoidal.

\begin{figure}
\centering
\includegraphics[width=0.5\textwidth]{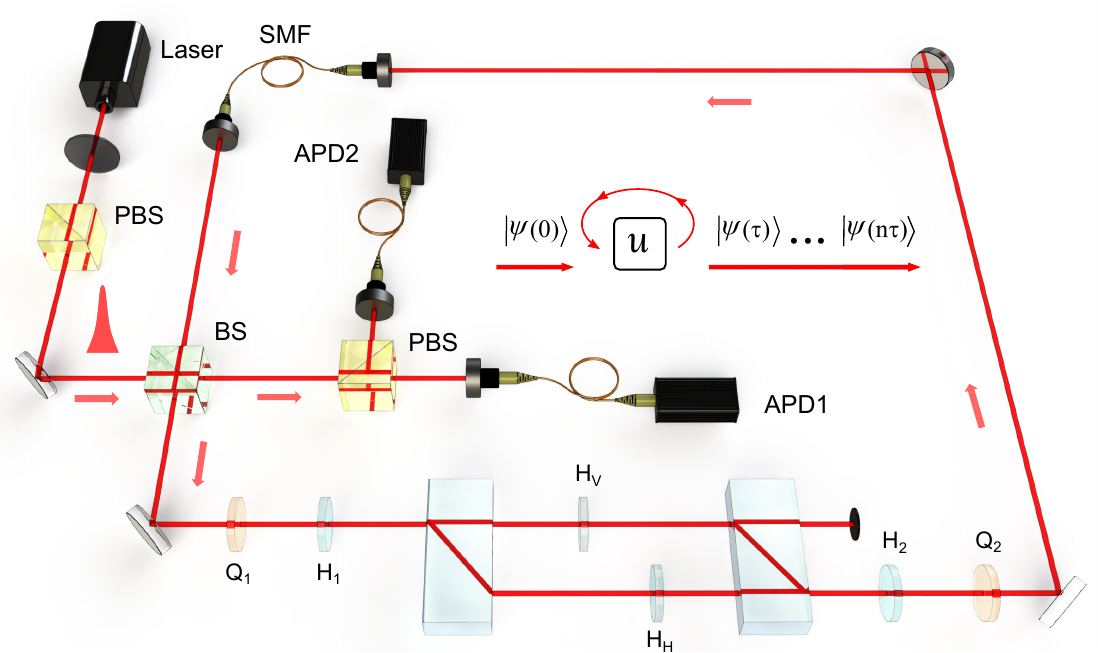}
\caption{Experimental schematic of the non-Hermitian cyclic-structure quantum evolution. A pulse laser is attenuated to the single-photon level via a set of neutral density filters and coupled into the cyclic quantum simulator through a beam splitter (BS). The non-unitary evolution operator $u$ is implemented by a series of optical elements. After each evolution $u$, about $10\%$ photons are reflected by the BS for projective measurement and the transmitted photons continue to go through the next cycle. The cyclic evolution process is depicted in the inset diagram. The state of the sensor is described as $|\psi(n\tau)\rangle=u(\tau)^n|\psi(0)\rangle$.}
\label{setup}
\end{figure}

{\it Experiment implementation.---}As a paradigmatic example, we experimentally implement the pseudo-Hermitian sensor described above using linear optics, dividing the process into many segments, each of which simulates a fixed time evolution governed by a pseudo-Hermitian Hamiltonian. The process can be extended to a real-time evolution and is regarded as a genuine quantum simulation in linear optics~\cite{XHS+14}.

In our experiment, we encode the qubit basis in the horizontal and vertical polarization states of photons, i.e., $|0\rangle\equiv|H\rangle$ and $|1\rangle\equiv|V\rangle$. As illustrated in Fig.~\ref{setup}, our experimental setup includes three stages: state initialization, time evolution, and projective measurement. The wave packets of photons are generated by a pulsed laser source with a central wavelength of $808$nm, a pulse width of $88$ps, and a repetition rate of $31.25$kHz. The pulses are attenuated to the single-photon level using neutral density filters. The photons are projected into the initial state $\ket{H}$ via a polarizing beam splitter (PBS) and then coupled in and out of an interferometric network through a low-reflectivity beam splitter (BS, reflectivity $10\%$), for time evolutions under the non-Hermitian Hamiltonian. Each segment of the time evolution is realized by the interferometric network involving several wave plates and two beam displacers (BDs). Such an interferometric network is capable of realizing an arbitrary $2\times2$ non-unitary operator~\cite{TRS18,XWZ+19,WXB+21}.

We implement a stroboscopic simulation of the non-Hermitian dynamics, partitioning the overall evolution time $t$ into $n$ short segments $t=n\tau$. In our experiment, we choose $a=1$ and $n=5$. The non-unitary time evolution operator in each segment is
\begin{equation}
u(\tau)=e^{-i (H-\mathcal{E}\mathbb{I}) \tau}=\begin{pmatrix}{\cos \alpha \tau}&{-i\frac{ \alpha}{\beta}\sin \alpha \tau}\\{-i\frac{ \beta }{ \alpha }\sin \alpha \tau} & {\cos \alpha \tau}\end{pmatrix},
\end{equation}
where $\alpha=\sqrt{(\mathcal{E}\delta+\lambda)(\mathcal{E}+\delta \lambda)/\delta}$ and $\beta=\mathcal{E}\delta+\lambda$. For the implementation of $u$, we further decompose it according to
\begin{align}
u=R_2\left(\theta_2,\phi_2\right)L\left(\theta_{H},\theta_{V}\right)R_1\left(\theta_1,\phi_1\right),
\end{align}
where the rotation operator $R_i(\theta_i,\phi_i)$ ($i=1,2$) is realized by a quarter-wave plate (QWP) at $\phi_i$ and a half-wave plate (HWP) at $\theta_i$, and the polarization-dependent loss operator $L$ is realized by a combination of two BDs and a pair of HWPs at $\theta_H$ and $\theta_V$. By adjusting the parameters of wave plates, we can simulate dynamics under Hamiltonians with tunable non-Hermiticity.

Each segment of the time evolution $u(\tau)$ is repeated through an optical cycle, in which photons propagate both in free space and through single-mode fibers. At the end of each cycle, the photons are re-coupled into the interferometric network through a low-reflectivity BS. The transmitted photons recycle for the next segment, while about $10\%$ of the photons are reflected and coupled out of the cycle and into the measurement module.

Finally, after photons have completed multiple cycles and are coupled out of the network by the BS, they register clicks at an avalanche photo-diode (APD) with a time jitter of $350$ps for detection. The population of photons is obtained through projective measurement, which consists of a PBS and APDs. The counts of the horizontally polarized photons $N_H$ and vertically polarized ones $N_V$ are registered in the measurement stage and are used to determine susceptibilities in our experiment~\cite{supp}. 

\begin{figure*}
\includegraphics[width=\textwidth]{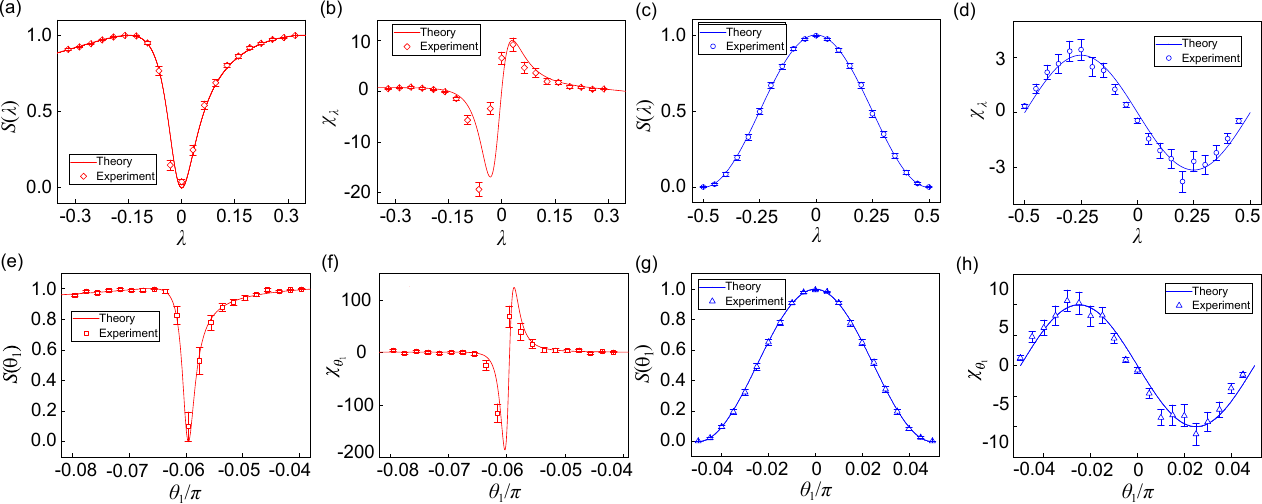}
\caption{Experimental results of the susceptibility of non-Hermitian and Hermitian sensors. The normalized population $S(\lambda)$ of the state $|H\rangle$ and the susceptibility $\chi_{\lambda}$ as a function of $\lambda$ for the non-Hermitian sensor (a)-(b) and its Hermitian counterpart (c)-(d), respectively. The normalized population $S(\theta_1)$ and the susceptibility $\chi_{\theta_1}$ as a function of $\theta_1$ for the non-Hermitian sensor (e)-(f) and its Hermitian counterpart (g)-(h), respectively. Symbols represent the experimental data and lines are the corresponding theoretical predictions with the initial state $|H\rangle$ and the evolution time $t=\pi/2\mathcal{E}$. Other parameters are $\delta=0.3$ and $\mathcal{E}=0.5$. Error bars are due to the statistical uncertainty in photon number counting.
}
\label{population}
\end{figure*}

{\it Measuring susceptibility.---}We construct the normalized population of state $|H\rangle$ through the measured photon counts $N_H$ and $N_V$, with
\begin{equation}
S=\frac{N_H}{N_H+N_V}.
\label{population_ex}
\end{equation}
The populations of the state $|H\rangle$ after $5$ segments of the time evolution governed by non-Hermitian and Hermitian Hamiltonians with the initial state $|H\rangle$ are shown in Figs.~\ref{population}(a) and ~\ref{population}(c), respectively. In the non-Hermitian case, the population changes more sharply compared to its Hermitian counterpart, with respect to small variations in $\lambda$, especially when $\lambda\in\left[-0.105,0.138\right]$.

In Figs.~\ref{population}(b) and ~\ref{population}(d), we show the measured susceptibility $\chi_{\lambda}$ as a function of $\lambda$ for both the non-Hermitian and Hermitian sensors. Since we take discretized values of $\lambda$ in our experiment, the susceptibility is numerically estimated according to

\begin{align}
\chi_{\lambda_i}=\frac{ S_{i+1}-S_{i}}{\lambda_{i+1}-\lambda_{i}},
\end{align}

where the subscript $i$ represents the data index. As illustrated in Fig.~\ref{population}(b), for the non-Hermitian sensor, we have the measured susceptibility $|\chi_{\lambda}|\leq 19\pm1$. Whereas for the Hermitian case in Fig.~\ref{population}(d), the measured susceptibility $|\chi_{\lambda}|\leq 3.8\pm0.6$. Therefore, the non-Hermitian sensor demonstrates a remarkable advantage for the estimation of $\lambda$.

{\it An application.---}In the experiment above, the external field $V$ is simulated by a set of linear optical elements, and the adjustment of the parameter $\lambda$ is realized by tuning the setting angles of the combination of wave plates (H$_1$, H$_2$ and H$_V$). As a primordial demonstration of non-Hermitian sensing, we now encode the signal $\lambda$ in the setting angle of a single wave plate. Our goal is to estimate the parameter $\theta_1$ of the wave plate $H_1$ with the non-Hermitian sensor. For this purpose, photons are initialized in the state $|\psi(0)\rangle=|H\rangle$. We fix the setting angles of the other wave plates as $\varphi_1=0.5\pi$, $\varphi_2=-0.5\pi$, $\theta_2=0.03\pi$, $\theta_V=0.43\pi$, which are determined by numerical simulations and optimization analysis. The populations of the state $|H\rangle$ after $5$ segments of the time evolution governed by the non-Hermitian Hamiltonian with the initial state $|H\rangle$ are shown in Fig.~\ref{population}(e). Apparently, the measured population becomes more sensitive to $\theta_1$ round $\theta_1=-0.06\pi$. The susceptibility $\chi_{\theta_1}$ is calculated from the measured population and shown in Fig.~\ref{population}(f). The maximum value of the susceptibility is $115\pm18$.

For comparison, we use a Hermitian sensor for the same task. We remove the BDs along with $H_2$, $H_H$, and $H_V$, leaving wave plates $Q_2(\phi_2)$-$H_1(\theta_1)$-$Q_1(\phi_1)$. We fix the setting angles of the wave plates to be the same as those of the non-Hermitian sensor, with $\varphi_1=0.5\pi$, $\varphi_2=-0.5\pi$. Figures~\ref{population}(g) and ~\ref{population}(h) show the population and the susceptibility of the Hermitian sensor. The measured susceptibility is $|\chi_{\theta_1}|\leq11\pm2$. Hence, the non-Hermitian enhancement ratio is $\sim 10.5$.

\textit{Non-Hermitian sensing under noise.---}Noise is a fundamental challenge for sensors as the performance of a sensor is ultimately limited by noise.
The non-Hermitian sensors demonstrated here outperform the Hermitian ones, particularly when the technical noises in experiments cannot be averaged out through repetitive measurements. Paradigmatic scenarios are sensing processes with strong background detection noises, or systematic errors; the latter usually results from certain unknown consistent deviations of the real experimental setting from the {\it a priori} scenario expected by the measurement agent~\cite{supp}. In the following, we take background noises~\cite{KG14,XZX+21} as an example, and experimentally confirm the non-Hermitian advantage.
%
For simplicity, we focus on the background noise due to the imperfection of detectors. The measured population of the state $\ket{H}$ in Eq.~(\ref{population_ex}) is then modified as

\begin{align}
S'=&\frac{N_H+N'_H}{N_H+N'_H+N_V+N'_V}\nonumber\\
\approx&S+(1-S)\frac{N_H'}{N}-S\frac{N_V'}{N}.
\label{population_niose}
\end{align}
Here $N'_{H,V}$  quantify the additional noisy photons in the corresponding polarization states caused by the background noise, and $N=N_H+N_V$.
In general, the background noise is completely random. The corresponding photon numbers are assumed to be  uniformly distributed, with $N_H'\in[0,\eta_HN]$ and $N_V'\in[0,\eta_VN]$, respectively.
With saturated photon detectors~\cite{HBL+17, XLD+20}, one can increase $N$ for better measurement precision, by considering a larger interrogation time, which typically involves more background photons as well. Hence, we assume fixed values of $\eta_{H,V}$ in the following.
Applying the error propagation  procedure on Eq.~(\ref{population_niose}), the measurement noise of $S'$ is
\begin{align}
 \Delta S'^2=&(\frac{\partial S'}{\partial S}\Delta S)^2+(\frac{\partial S'}{\partial N_H'/N}\Delta \frac{N_H'}{N})^2+(\frac{\partial S'}{\partial N_V'/N}\Delta \frac{N_V'}{N})^2\nonumber\\
 =&(1-\frac{\eta_H+\eta_V}{2})^2\Delta S^2+\frac{1}{12}[(1-S)^2\eta_H^2+S^2\eta_V^2].
 \label{ds}
\end{align}
The first term on the right-hand side of Eq.~(\ref{ds}) (in the second line) comes from the fluctuation of photons, which can be averaged out through repetitive measurements. Namely, the term tends to zero when $N\gg 1$. By contrast, the second term comes from the background noise, which is independent of $N$ and cannot be averaged out. 

\begin{figure}
\includegraphics[width=0.5\textwidth]{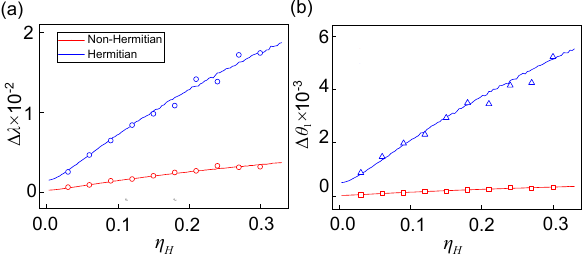}
\caption{Experimental results of the measurement uncertainty. Measured standard deviation $\Delta \lambda$ in (a) and $\Delta \theta_1$ in (b) as a function of the strength $\eta_{H}$ of the background noise. Symbols representing the results are obtained by repeating measurements for $50$ times for each of the noise strengths chosen in our experiment. The lines are numerical simulations that are obtained from $10000$ measurements for each noise strength.}
\label{noise}
\end{figure}

We experimentally demonstrate the performance of the non-Hermitian sensor in the presence of background noise. 
We use two light-emitting diode (LED) which are placed close to the APDs, to simulate the background noise. We adjust the intensity of the LEDs to ensure that the changes in $N'_{H,V}$ are proportional to $N$. For simplicity, we choose $\eta_V=1.2\eta_H$. We choose the working point of the sensor, i.e., $\lambda$ corresponding to $|\chi_{\lambda}|_\text{max}$, and repeat measurements for $50$ times for each of the noise strengths chosen in our experiment. For better contrast, we adopt the same integration time and photon counts for the non-Hermitian and Hermitian sensors, to simulate the saturation for detectors. Although the input photon intensity in the non-Hermitian case is higher than that in the Hermitian case, it should not matter for saturated detectors, as the measurement results would be the same when further increasing the input photon intensity of the Hermitian detector.
For the measured population of the state $\ket{H}$ with the measurement noise $\Delta S'$ in Eq.~(\ref{ds}), the measurement uncertainty of the parameter $\lambda$, namely the minimum error between the estimated and true values of $\lambda$, is quantified by $\Delta \lambda\approx \Delta S'/|\chi_{\lambda}|_\text{max}$ through the propagation of error ~\cite{supp}.

In Fig.~\ref{noise}, we show how the measurement uncertainty varies as a function of the noise strength $\eta_H$, for both the non-Hermitian and Hermitian sensors. The results show that, with increasing noise strength, the measurement uncertainty $\Delta \lambda$ ($\Delta \theta_1$) deteriorates in both cases. However, $\Delta \lambda$ ($\Delta \theta_1$) of the non-Hermitian sensor is always smaller compared to the Hermitian one. Thus, the non-Hermitian sensor possesses improved performance compared to the Hermitian counterpart in the presence of background noise. As such, non-Hermiticity allows one to reduce the measurement uncertainty, over a conventional Hermitian sensor, both with and without noise.

\textit{Discussions.---} We propose and demonstrate a generic non-Hermitian sensing scheme that does not rely on the presence of EPs.
Simulating the non-Hermitian dynamics of the sensor-field system, we show that the non-Hermitian sensor is superior to its Hermitian counterpart, both in terms of susceptibility, and in terms of performance against technical noises.
While the non-Hermitian sensor in our single-photon experiment is time-independent, it is possible to achieve time dependence through electro-optical modulators, which should open up fresh opportunities for novel sensor design.
Our scheme exploiting a general non-Hermitian Hamiltonian is also feasible in other physical systems, such as the solid-state spins~\cite{WLG+19}, superconducting transmon circuits, and single trapped ions~\cite{DSZ+21, NAJ+19}, where non-Hermitian physics are routinely simulated on the single-qubit basis and beyond. Furthermore, while our demonstration is performed on the single-qubit level, it is possible to extend the sensing scheme to multi-qubit cases. Our work thus paves the way toward a new class of non-Hermitian sensors.

\begin{acknowledgments}
This work is supported by the National Key R\&D Program of China (Grant Nos. 2023YFA1406701 and 2023YFA1406703), National Natural Science Foundation of China (Grant Nos. 12025401, 92265209, 12304572, 12374479, and 12104036) and the fellowship of China Postdoctoral Science Foundation (Grant No. 2022M721256).
\end{acknowledgments}







\end{document}